\documentclass{article}
\usepackage{PRIMEarxiv}

\usepackage[utf8]{inputenc} 
\usepackage[T1]{fontenc}    
\usepackage{hyperref}       
\usepackage{url}            
\usepackage{booktabs}       
\usepackage{fancyhdr}       
\usepackage{graphicx}       

\begin{document}

\title{Chameleon: A Semi-AutoML framework targeting quick and scalable development and deployment of production-ready ML systems for SMEs}

\author{
  Johannes Otterbach, Thomas Wollmann \\
  Merantix Labs GmbH \\
  Berlin\\
  \texttt{\{johannes.otterbach, thomas.wollmann\}@merantix.com} \\
}

\maketitle

\begin{abstract}
Developing, scaling, and deploying modern Machine Learning solutions remains challenging for small- and middle-sized enterprises (SMEs). This is due to a high entry barrier of building and maintaining a dedicated IT team as well as the difficulties of real-world data (RWD) compared to standard benchmark data.
To address this challenge, we discuss the implementation and concepts of Chameleon, a semi-AutoML framework. The goal of Chameleon is fast and scalable development and deployment of production-ready machine learning systems into the workflow of SMEs. We first discuss the RWD challenges faced by SMEs. After, we outline the central part of the framework which is a model and loss-function zoo with RWD-relevant defaults. Subsequently, we present how one can use a templatable framework in order to automate the experiment iteration cycle, as well as close the gap between development and deployment. Finally, we touch on our testing framework component allowing us to investigate common model failure modes and support best practices of model deployment governance.
\end{abstract}

\keywords{Semi-AutoML \and production systems \and IT infrastructure \and ML development and deployment \and computer vision}

\section{Introduction}

Modern Machine Learning and Artificial Intelligence systems hold the promise to disrupt many workflows of small- and medium-sized enterprises (SMEs) and propel them forward in the digitization efforts. This will result in more efficient use of resources as well as more targeted fulfillment of customer demands. Despite the promise of the technology, the development and deployment of state-of-the-art Machine Learning systems remains challenging for SMEs: If Machine Learning projects are being tackled at all, they are often terminally stuck in the proof-of-concept (PoC) stage as scaling and productionizing them requires a team of dedicated software- and data-engineers as well as data scientists. As a consequence many SMEs are cut-off from the technological progress, reserving their use to big companies and thus inhibiting their competitiveness \cite{oecd-report}. In addition to this, SMEs adopting external AI systems incur risks due to technical debt, new attack surfaces and reliability of the technology regarding model generalization and robustness, adversarial attacks, dataset bias, and other complications in connection with Machine Learning systems \cite{ai-safety-concrete, technicalDebt-ml, reliable-ml}.

\begin{figure}[t]
  \centering
  \includegraphics[width=.8\textwidth]{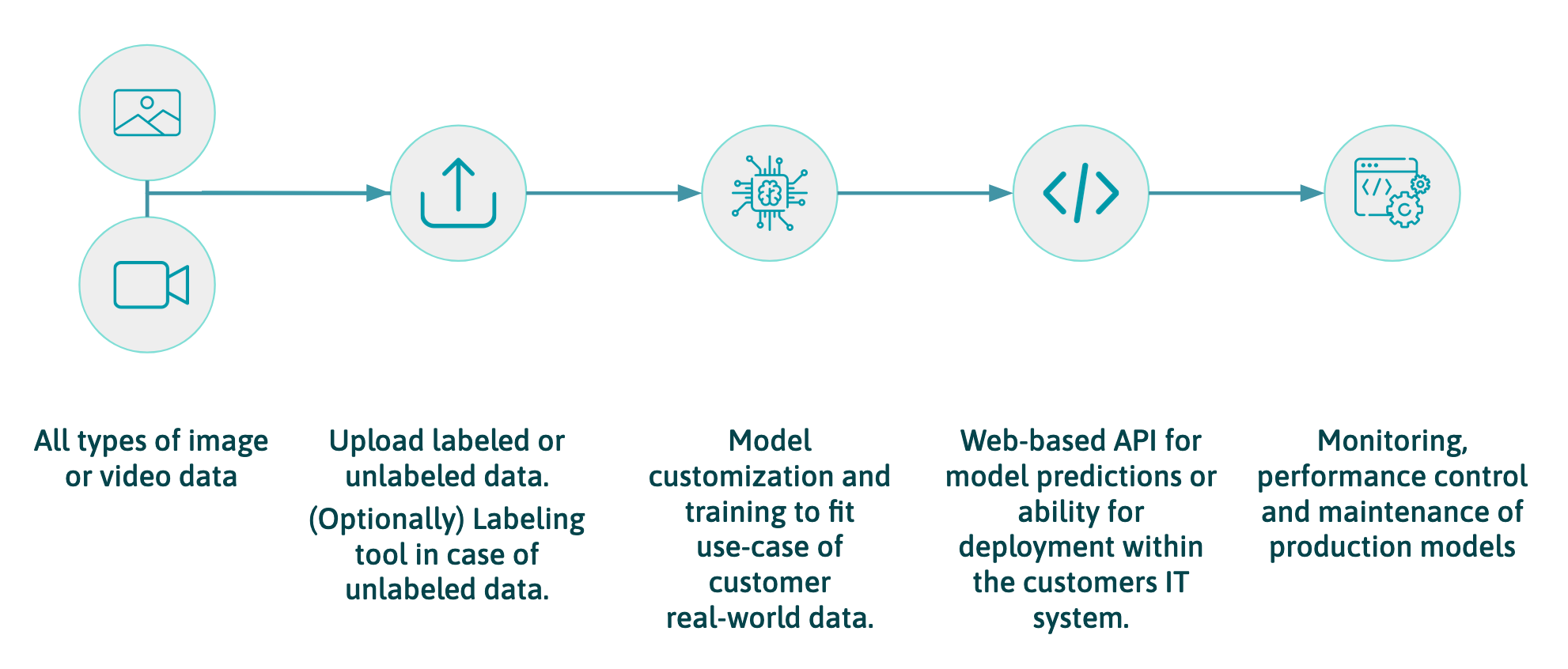}
  \caption{Functional diagram of the Chameleon semi-AutoML framework. The framework is generally applicable to any data domain and provides optional labeling tools. It then ingests and prepares the data for model training with minimal effort. The resulting model is ready to be deployed and monitored without further tweaks to the environment.}
  \label{fig:merantix_chameleon}
\end{figure}

In this paper we present Chameleon, a semi-AutoML \cite{AutoML} software framework developed at Merantix Labs, to address these challenges. The framework allows to tackle the diversity in workflow and use-cases of SMEs in a highly scalable manner by providing tools to efficiently orchestrate a big training infrastructure, abstracting common use-cases, deploying monitoring tools, and finally scale and virtualize the production deployment for inference. In addition to this the framework provides an optional labeling tool and incorporates standard testing procedures to assess the vulnerability of the Machine Learning system to common problems outlined above (see Fig. \ref{fig:merantix_chameleon}).

As a result of this, the cost of developing and deploying a robust Machine Learning model into production is significantly reduced and the burden of hiring and maintaining a dedicated Machine Learning-engineering team can be outsourced from the SMEs to specialized companies.

While the Chameleon framework is very flexible and can be applied to any data-modality, we focused on the development of computer vision use-cases such as image classification, object detection, image and instance segmentation.

The report is structured as follows: First we will discuss the different challenges of real-world data (RWD) that SMEs face and show how they differ from standard benchmarks available to academic research. We then dive into the overall architecture of Chameleon that allows us to quickly iterate on customer projects and finally show how the testing framework is integrated into the platform.

\section{Data Challenges for SMEs in Computer Vision Use-Cases}
Computer vision is a central and maybe the oldest field of study in the modern Deep Learning era \cite{Imagenet2015}. It fielded the recent breakthrough of Artificial Intelligence techniques with the Imagenet challenge in 2012 \cite{Imagenet2015, AlexNet2012} and heralded a decade of innovations ranging from natural language processing, computer vision, and video processing to speech recognition and optimization using techniques from (un-)supervised and generative modeling to reinforcement learning.

However, the development of these techniques heavily relies on the availability of open-source datasets that serve as benchmarks metrics against which to test new algorithms. Within the computer vision domain, such datasets are, Imagenet \cite{Imagenet2015}, PascalVOC \cite{Everingham10},  Flicker 100M \cite{yfcc100m}, CoCo \cite{msft-coco}, Cityscapes \cite{cityscapes}, etc. As a general rule of thumb public datasets are assembled from common data modalities on the internet. As such they do not represent distributions from highly specialized domains such as medical image analysis \cite{grand-challenge}, remote sensing or visual quality control. Moreover, pre-training on these public datasets does not necessarily transfer to these specialized domains \cite{liu2017detecting} diminishing the utility of the public datasets. Consequently, it is important to understand the differences between these benchmark datasets and the data found at SMEs. 

\begin{figure}[t]
    \centering
    \includegraphics[width=.355\textheight]{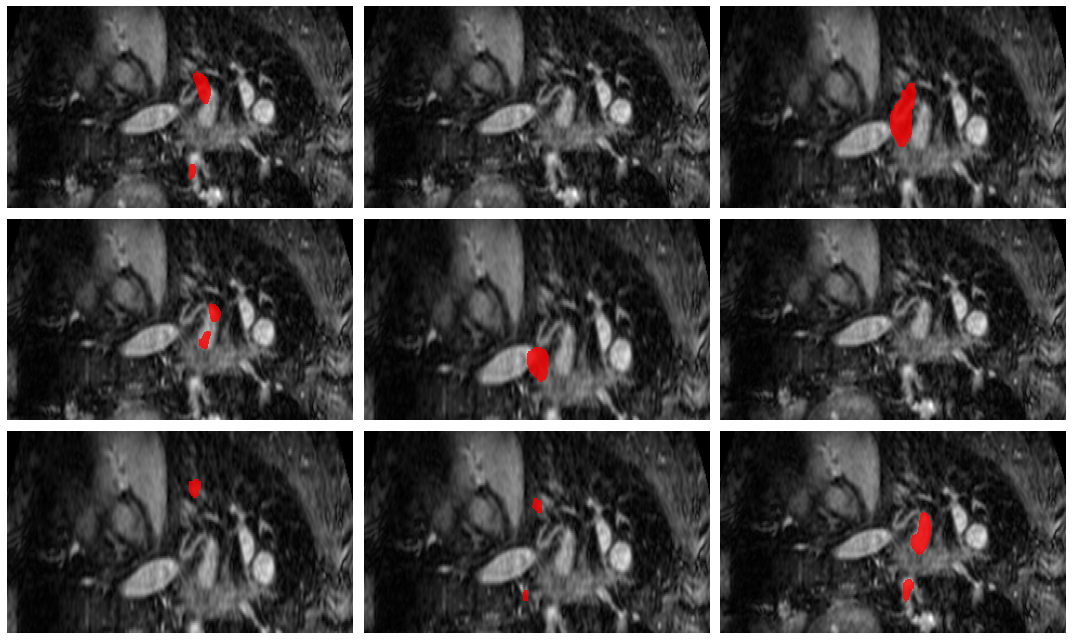}
    \hspace{.2cm}
    \includegraphics[width=.28\textheight]{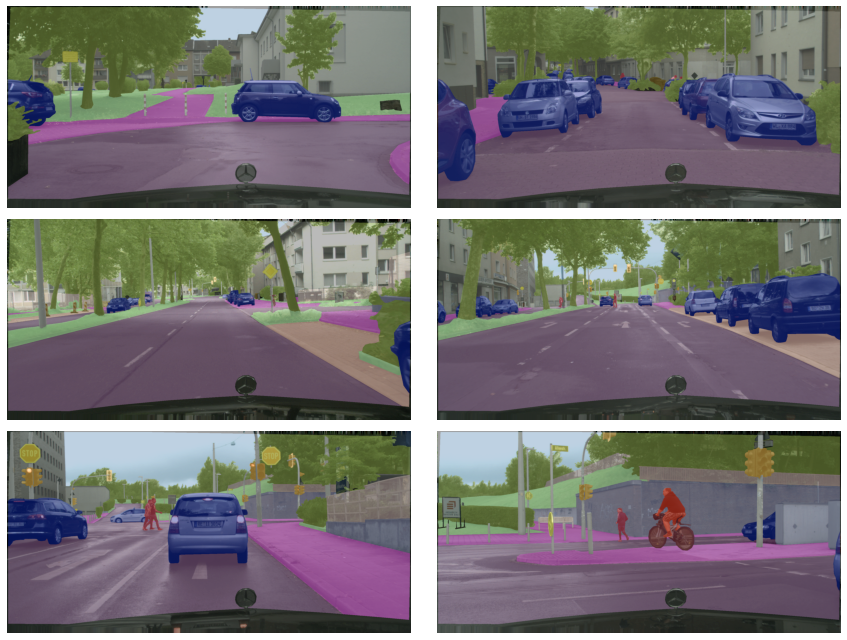}
    \caption{(\textit{left}) Prostate segmentation examples taken from \cite{medical-decathlon}, (\textit{right}) Cityscapes segmentation samples taken from \cite{cityscapes}. While both task are segmentation tasks from a technical perspective, the label density is significantly higher and more balanced in the Cityscapes dataset, while the label density is low in the prostate segmentation task, but the data is much more regular due to similar poses. Evaluating on the standard benchmark of Cityscapes hence does not necessarily make predictions of algorithmic performance on different data such as medical segmentation.}
    \label{fig:segmentation_data}
\end{figure}

For example X-ray, Magnetic Resonance Imaging (MRI) or visual quality control (VQC) data is typically more structured as they are acquired in a controlled environment and tend to have the same position of body or tool parts in the image as opposed to natural images. In addition the data is not always obtained using natural image techniques but are false-color images generated through post-processing of frequency and intensity data of specialized equipment such as Light Detection and Ranging (LIDAR), MRI or fluorescence microscopy \cite{wollmann201770}. These dataset biases often allow for more efficient modeling approaches \cite{Maier_2019} and require different data-preprocessing techniques compared to the ones developed in the general Machine Learning community.

A second example within the VQC domain is the detection of small defects. Typically this use-case is formulated as an image segmentation task, where we do not only require the prediction if an image contains a defect, but in addition, also where the defect is located within the image. Since by nature these defects are small we face a strong class imbalance problem where most of the segmentation is the no-defect ground truth, while we only have a few defect pixels, which often are also broken up into several classes. Similar types of problems also exist in the medical image domain. This stands in contrast to public image segmentation benchmark datasets such as Cityscapes or CoCo, which are developed for natural scene understanding and display significantly reduced class imbalance (see Fig. \ref{fig:segmentation_data}). In addition the strong imbalance and small defect regions lead to pronounced instabilities in standard segmentation losses \cite{reinke2021common}.

As a result of these differences, architectures, loss functions, training curricula and other techniques reaching state-of-the-art (SOTA) performance on standard benchmarks do not always transfer to RWD use-cases. To quickly assess the transferability of new developments to datasets of SMEs, Chameleon incorporates a model zoo of different computer vision architectures and loss functions. As outlined below, those components are templatable, allowing for fast iteration and experimentation to reduce time to deployment of performant models. Moreover, these components come with relevant defaults that are reflective of the dataset difference between the benchmark and RWD domains.

\section{Chameleon framework}
The IT industry has seen an increasing number of frameworks for developing Machine Learning models, such as PyTorch \cite{pytorch}, Tensorflow \cite{tensorflow}, Keras \cite{keras}, JAX \cite{jax} for Python, Owl \cite{owl} in OCaml, Flux \cite{flux} for Julia, flashlight \cite{flashlight} for C++. However, the field is still lacking tools for the end-to-end development of Deep Learning pipelines, similar in spirit to packages such as scikit-learn \cite{scikit-learn} or Spark \cite{spark}, and many SOTA applications are only available in form of Jupyter Notebooks \cite{jupyter} in more-or-less maintained open-source projects. One of the reasons for the lack of end-to-end solutions is the significantly larger infrastructure overhead of training Deep Learning models, as they often require accelerators, e.g. GPUs, cloud-storage for large datasets in the range of several GB up to PB (at the point of writing this report), monitoring tools as well as inference deployment and integration. The result is a large gap in the ability to deploy a working proof-of-concept model into production \cite{scully2015}. 

We address this shortcoming by building a semi-AutoML pipeline that can be orchestrated and deployed with minimal efforts. The core of the pipeline is formed by Kubernetes to deploy containers into cloud-based solution such as the Google cloud, AWS or Azure. The containers are templated through a setup and are built during the Continuous Integration and Deployment (CI/CD) step. Once this setup is in place the deployment of Machine Learning systems becomes much easier as we can break down the steps of a development cycle into individual container deployments and the CI step ensures stability and reproducability of the development \cite{graetz}.

We chose to separate the Extract-Transform-Load (ETL) stages from the training and inference stage of the pipeline. The ETL stage comprises the data-preprocessing steps and allows quick and repeated experimentation with different data-augmentation and data-enhancement approaches, necessary to address the RWD challenges outlined in the previous section. In addition it allows to built reusable components for ETL that can in many cases be reused end-to-end for new customer use cases. To facilitate fast iteration of model developments, we equipped the core Machine Learning module with several additional abstractions: Loss functions and model types as well as training loops and inference code-paths. In this way the training and inference can be orchestrated through a Terraform template. In the ideal scenario, this results in minimal code that needs to be developed to integrate a new data source into the pipeline and moves the development of high-performance Machine Learning systems closer to low-code environments. In addition to these benefits, the templating lets us auto-generate good pipelines based on well-chosen and field-tested heuristics. In order to address more dynamic data environments, where data can become stale over time, we can integrate the models into a CI/CD cycle to regularly roll out updates to the models without performance regressions.

\section{Testing Framework}
\begin{figure}[t]
  \centering
  \includegraphics[width=.98\textwidth]{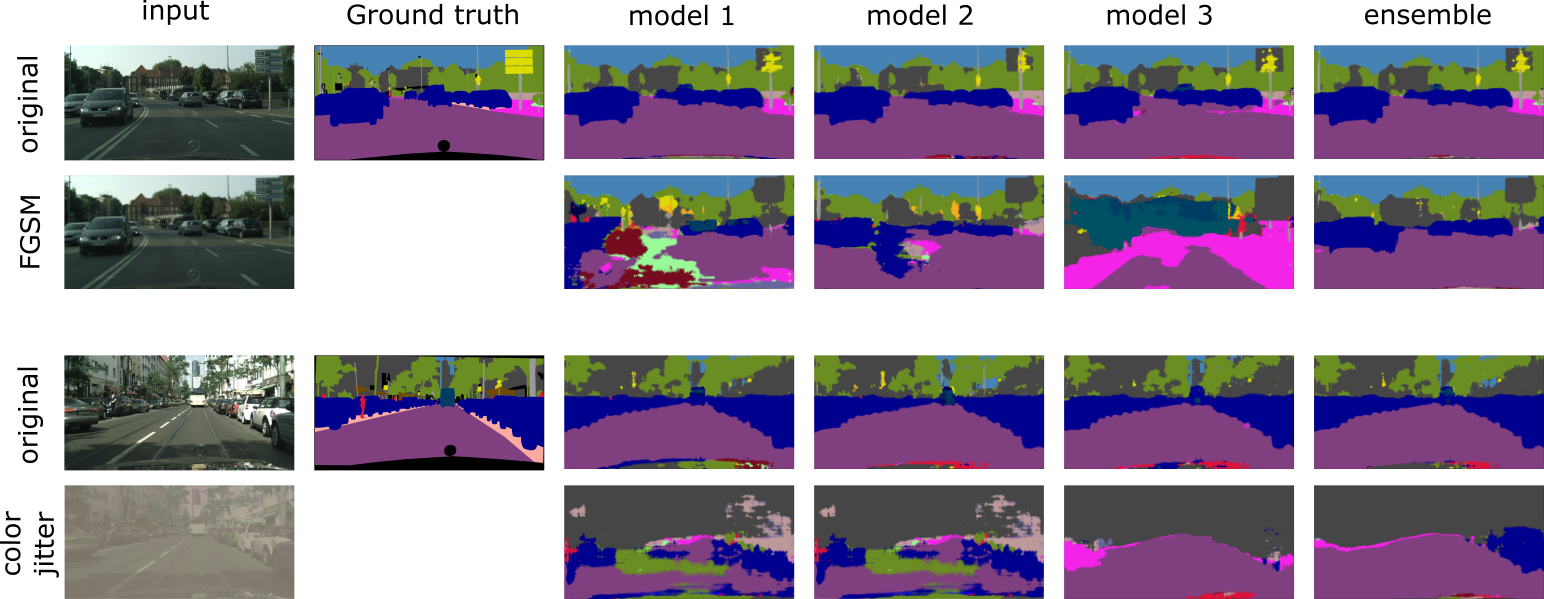}
  \caption{Gradient-based and color jitter adversarial attack against image segmentation models. Each individual model is strongly fooled by the adversary as can be seen by the predicted label map of the outputs. However, using an ensemble of the models results in a significantly more robust prediction making it harder for the attacker to fool the system. The testing framework of Chameleon facilitates such investigations at a much faster rate.}
  \label{fig:adversarial_segmentation_attack}
\end{figure}
Responsible deployments of Machine Learning pipelines into production require following a reliable and responsible model governance process \cite{model-cards, ai-safety}. A central part of this process is the investigation of generalization, biases and robustness to adversarial attacks. This is often a manual process since it requires experimentation with the trained model and subjecting it to a variety of different scenarios. Using Chameleon, we automate many pieces of this process in order to reduce the manual labour and speed up the investigation process. For example we implement data-augmentation to simulate of different image conditions, such as fog or blur to simulate common perturbations and corruptions \cite{hendrycks2019robustness}. We also support standard white-box adversarial attacks, such as Fast-Gradient-Sign-Method (FGSM) \cite{goodfellow2014explaining}, iterative-FGSM \cite{kurakin2017adversarial} and others, on segmentation data \cite{xie2017adversarial, fischer2017adversarial}. Finally we evaluate the model performance under these perturbations and visualize the results (see Fig. \ref{fig:adversarial_segmentation_attack}). This process speeds up the analyst in assessing the model's capabilities and improves modeling as it surfaces model failure modes early, allowing to develop various mitigation strategies throughout the full pipeline such as model ensembling for segmentation mask predictions.
In addition, the inference path in the Chameleon framework along with the ETL pipeline lets us easily investigate the generalization of the model to other datasets within a relevant domain or even outside the domain. Doing this highlights complementary failure modes and informs the SMEs about limitations of the model so they can be monitored in production.

\section{Summary}
SMEs face a large gap between developing PoCs of modern Machine Learning systems and their deployment into production cycles. Not only is developing SOTA models expensive in terms of compute as well as human resource, but SMEs also face a challenge in modeling due to large differences between their own data and standard benchmarks of the Machine Learning community. To close this gap, we developed the Chameleon framework which allows us to quickly integrate and iterate on modeling real-world data using a zoo of models and training procedures. Due to our templatable and containerized semi-AutoML framework, we can quickly deploy models into a production workflow, reducing the integration burden of customers and saving them from hiring and maintaining a dedicated Machine Learning team. We incorporate an automated testing framework to assess common Machine Learning model failure modes, such as adversarial attacks and image defects, thus ensuring best practices for model deployment governance and increasing trust in the capabilities of automated decision engines.

\section{Acknowledgements}
The research leading to these results is funded by EU ECSEL Project \textit{SECREDAS Cyber Security for Cross Domain Reliable Dependable Automated Systems} (Grant Number: 783119).

\bibliographystyle{unsrt}

\end{document}